**Visual Conceptualizations and Models of Science**

*Katy Börner, Cyberinfrastructure for Network Science Center, School of Library and Information Science, Indiana University, 10th Street & Jordan Avenue, Wells Library 021, Bloomington, IN 47405, USA*
katy@indiana.edu

*Andrea Scharnhorst, The Virtual Knowledge Studio for the Humanities and Social Sciences - VKS*
*Royal Netherlands Academy of Arts and Sciences, Cruquiusweg 31, 1019 AT Amsterdam, The Netherlands*
andrea.scharnhorst@vks.knaw.nl

This *Journal of Informetrics* special issue aims to improve our understanding of the structure and dynamics of science by reviewing and advancing existing conceptualizations and models of scholarly activity. Several of these conceptualizations and models have visual manifestations supporting the combination and comparison of theories and approaches developed in different disciplines of science.

The term "model" and "conceptualization" has diverse definitions in different disciplines and use contexts. In this issue, 'conceptualization' refers to an unifying mental framework that identifies the boundaries of the system or object under study, its basic building blocks, interactions among building blocks, basic mechanisms of growth and change, and existing laws (static and dynamic). The term 'model' refers to a precise description of a system or object under study in a formal language, e.g., using mathematical equations or computational algorithms. An older more comprehensive definition of 'science' using a description by Cohen "… in its oldest and widest sense the term science (like the German word *Wissenschaft*) denotes all ordered and reliable knowledge—so that a philologist or a critical historian can truly be called scientific …" (Cohen, 1933). Consequently, 'science of science' refers to the scientific study of all scholarly activities in the natural sciences, the social sciences, and the arts and humanities. Given the number of scientific disciplines involved in the study of all sciences, the phrase 'sciences of sciences' might be more appropriate yet the singular versions are commonly used.

Subsequently, we discuss challenges towards a theoretically grounded and practically useful science of science and provide a brief chronological review of relevant work. Then, we exemplarily present three conceptualizations of science that attempt to provide frameworks for the comparison and combination of existing approaches, theories, laws, and measurements.

Finally, we discuss the contributions of and interlinkages among the eight papers included in this issue. Each paper makes a unique contribution towards conceptualizations and models of science and roots this contribution in a review and comparison with existing work.

To strengthen the historical embedding of current and anticipated future work, this special issue starts with an invited contribution by Eugene Garfield on the history of 'science of science' studies. Eugene Garfield's paper combines personal autobiographic memories as an eye-witness from the very beginning of scientometrics with data driven historiographic citation analysis. This way, the paper makes a convincing argument for the combination of qualitative and quantitative approaches and the communication of results by visual means in order to exploit the full power of many different sciences to study science itself.

**Science of Science Challenges**

A true science of science will have to be theoretically grounded and practically useful. It will support repeatability, economy, mensuration, heuristics, and consilience (Wilson, 1998), a view also emphasized by Henry Small (Small, 1998). It will increase the reliability of beliefs and assumptions by eliminating or minimizing errors and illusion which obstruct human knowledge while cultivating rather than suppressing doubt.

It will combine quantitative approaches (automatic, data driven, providing large-scale but coarse context) and qualitative approaches (manual, interview and survey based, providing small-scale but detailed 'high resolution inserts') in the study of science as intended at the very beginning of this field (see the contribution by Eugene Garfield).

It will draw on and advance existing theories that come from many domains of science with vastly different cultures, approaches, and toolsets such as the philosophy of science; the sociology of science; the social-psychological perspective on individual scholars and groups; scientometrics, informetrics, and webometrics; the history of science; cultural studies of science; operational research on science, statistical analysis and mathematical modeling of science; the economics of science; and science policy. It will benefit from increased data digitization and computation that lead to major advances in webometrics (Ingwersen & Björneborn, 2004; Thelwall, 2004), virtual and large scale ethnography (Hine, 2000), and information visualization (Card, MacKinlay, & Shneiderman, 1999) among others.

Hence, it will be beneficial to identify and define major scientific challenges and practical applications, to find more efficient means to utilize and interlink relevant knowledge scattered across different fields, to communicate results at a detail that supports reimplementation/application by others, to validate approaches and models in a scientifically rigorous, repeatable way, and to make the best datasets, tools, and results available for research, education, and practice. Conceptualizations of science that draw on and extend the rich history of science of science studies can play a major role in addressing these challenges.

**Science of Science History**

The study of knowledge and meaning and how scientific knowledge is produced has a long tradition in philosophical as well as in historical research. With the growth of scholarly activities, the institutionalization of science, the emergence of the disciplinary structure of science[1], and the role of scientific knowledge during the industrial revolution the social function of science has been analyzed (Bernal, 1939). Outside the Anglo-American language space the emerging field of "science theory" or "science and technology studies" can be found under notions such as "Wissenschaftsforschung oder

---

[1] The growth of disciplines in science has been the object of philosophical and historical as well as scientometric studies. See Wouters 1999 for further references.

Wissenschaftstheorie"[2] (German), "wiedza o nauce" (Poland) [3], "Науковедение" (Russian), or "Wetenschapsdynamica" (Dutch) pointing to an investigation of all scholarly activities The "sciences" (in this broad sense) became another special social system to be studied and described in addition to economy, politics, or social behaviour. But, these investigations did not lead to the formation of a major discipline comparable to economics, political sciences, or sociology. Instead, different types of science studies are carried out more or less independently by scholars in different scientific areas. To provide a platform of integration and reference in this interdisciplinary discourse is one motivation for this special issue.

This editorial is not the place to write the history of the science of science nor is it the place to review existing histories – see Garfield's contribution in this issue and (Garfield, 2003; Wouters, 1999). But, we can state that after World War II the need to better understand the growing "science system" led to a number of academic activities (books, journals, conference series and scientific societies) which finally cumulated in an very heterogeneous research field. Examples of influential books are the already mentioned social theory of science of Bernal (1939), the sociological approach by Merton (1973), and the aspect of measuring science systematically (Price 1963; Nalimov, Mulchenko 1969). Also in this period, new journals came into existence; *American Documentation* was founded in 1950 (and became JASIS 1970 and JASIST in 2000); *Scientometrics* in 1978. Recently, Leydesdorff has analyzed the disciplinary position of the field looking at the citation environment of these journals (Leydesdorff, 2007). The *Society for Social Studies of Science (4S)* was founded in 1975, followed by the *European Association for the Study of Science and Technology (EASST)* in 1981. The *Science and Technology Indicators Conference Series* started in 1988 and the *International Society of Scientometrics and Informetrics (ISSI)* was founded in 1993.

The aim to better understand scholarly activity was triggered by:

- the need for a better understanding of a phenomenon penetrating all areas of society forming the basic of economic wealth, and
- the need of national states and of international institutions such as the OECD to monitor an area of substantive spending of public resources.

Science policy analysis became a regular part of the social studies of science with journals such as "Science and Public Policy"(founded 1974) and "Research Evaluation" (founded 1991; see also (Godin, 2005)). Links to other fields such as the older one of information and documentation as well as the just emerging computer sciences are worth mentioning. Last but not least, increasingly popular databases such as the *Science Citation Index* by the Institute for Scientific Information (founded 1960) influenced the thinking about science[4]. Books such as "Towards a Metric of Science" (Elkana, Lederberg, Merton, Thackray, &

---

[2] See as an example the engagement of Wilhelm Ostwald for an bibliographic institute " Die Brücke – Institut zur Organisation geistiger Arbeit" (1911) in ((Hapke, 2008).

[3] In the 1920's, a group of Polish philosophers and sociologists (Florian Znaniecki, Maria Ossowska, Stanislaw Ossowski and others) wrote a program of the modern theory of knowledge (Wouters, 1999).

Zuckerman, 1978) presented a combination of qualitative and quantitative analysis of science. Later on, quantitative studies and qualitative studies seemed to follow separated ways in the study of the sciences. This is also visible in the existence of two handbook series: The *Handbook of Science and Technology Studies* and the *Handbook of Quantitative Studies of Science and Technology*. The formation of the International Society for Scientometrics and Informetrics (ISSI) (main focus: quantitative) as an alternative to the already existing Society for Social Studies of Science (4S) is another visible example of this separation. To create a bridge between quantitative and qualitative studies of science is a second motivation for this issue.

Since the middle of the last century, the field of science studies has grown and differentiated. Different turning points have been stated as the *cognitive turn* (Callon, Courtial, Turner, & Bauin, 1983) and the *evaluative turn* in scientometrics (Narrin, 1976; van Leeuwen, 2004), the *ethnographic turn* in science studies (Knorr-Cetina, 1995), the *informational turn,* which questions the role of information sciences as data and or method provider (Thelwall, Wouters, & Fry, 2008), or the more general *communicative turn* in the social sciences which also influenced the study of science (Leydesdorff, 2008). In recent years, we also observe an encountering between information sciences rooted in computer sciences and information and library sciences, science studies, and scientometrics. Recent achievements in information science and computer science together with the increased availability of digital scholarly data and computing resources have made it possible to analyze, model, and visualize science at an unprecedented scale and with a high level of sophistication (Börner, Chen, & Boyack, 2003). Networks of publications and their citation patterns, word use, collaborating researchers, or topics in e-mail threads have been measured, visualized and analysed over time. With the emergence of "network science" as a new cross-disciplinary approach (Barabasi, 2002; Börner, Sanyal, & Vespignani, 2007; Huberman, 2001; National Research Council of the National Academies, 2005; Price, 1963; Watts & Strogatz, 1998; Watts, 2003) and in particular with the achievements of visualizing knowledge domains in the information sciences and singular events such as turning points or paradigmatic shifts (Chen, 2003; Mane & Börner, 2004; Shiffrin & Börner, 2004), old dreams of mapping the sciences (Marshakova, 1973; Small & Griffith, 1974) can now be realized. An example of this approach are the so-called "maps of science" which show all scientific disciplines (Boyack, Klavans, & Börner, 2005; Chen, 2006; Leydesdorff & Rafols, 2009). From this perspective one could add another - the *computational scientometrics turn* – a term coined by Lee Giles to refer to scientometrics studies that use terra-bytes and advanced computing infrastructures – and the *visualization turn* to this sequence of paradigmatic changes in studying science.

The new techniques revive old questions. The idea of an "ecology of science" re-emerges in the visualizations. In the map of science, social sciences and humanities are securely interlinked with the natural sciences via mathematics and other sciences – even so major journal publication databases cover only a small fraction of soft sciences such as sociology, psychology, and communication science. Another question concerns the source of new ideas: how much do they rely on persons versus agreed knowledge codified in communication? Improved techniques in data mining allow nowadays for identification of authors in large databases with a sufficient accuracy at least for statistical analysis. Tracing the authors we can build bridges to biographic, interview and field based studies of the authors (cf. (Gaughan & Ponomariov, 2008)). But, it now also seems to be possible to link between acts of communication, actors, and institutions in a more systematic way than was possible in the past. Large-scale data analysis and visualization does not take away the question of how to interpret the now visible structures. On the contrary, questions about the driving forces behind visible changes in scientific communication, about our understanding of scientific growth and development, and the original questions at the beginning of the field seem to return even more fundamentally. To contribute to a reflection about the concepts and theoretical assumptions hidden in empirical case studies and to the search of an <u>operationalization of theoretical concepts in terms of measurement and empirical evidence</u> is the third motivation for this issue.

**Science of Science Conceptualizations**

Many areas of science have developed elaborated frameworks to compare, combine, and/or communicate existing datasets, models, and insights. Examples are geographic reference systems that support the 'layering' of data on utility pipes, streets, property lines, etc. in geographic information systems; astronomical reference system that make it possible to retrieve any existing dataset or insight for a given segment of the sky; or multi-level frameworks in biology that support model integration over a wide range of physical scales.

Existing conceptualizations and models of science comprise philosophical concepts such as those proposed by (Bernal, 1939; Kuhn, 1962; Popper, 1979), (utopian) stories (Lem, 1997; Wells, 1938), visual drawings (Otlet, 1934), social and cognitive theories (Merton, 1973; Simon, 1996), empirical measurements (Garfield, 1979; Price, 1962), and mathematical theories (Goffman & Warren, 1969; Yablonskij, 1986). Ideally, they state the boundaries of the modeled system or object; basic building blocks, e.g., units of analysis or key actors; interactions of building blocks (e.g., via coupled networks); basic mechanisms of growth and change; and the underlying static and dynamic laws.

It is our belief that a shared conceptualization of science can provide the intellectual framework to interlink and puzzle together the many models in existence today leading to a more comprehensive description and understanding of the structure and dynamics of science. Ideally, such a framework would provide a visual depiction of the 'landscape' of existing and missing models.

Subsequently, we discuss three exemplary conceptualizations of science. The first captures science as manifested in scientific papers, the second characterizes science as accumulation of knowledge, and the third sees science as a search in an abstract problem space. Note that it is beyond the scope of this editorial to present a full description of each conceptualization. Instead, we hope to communicate the strength and limitations of each approach and to inspire alternative conceptualizations and comparisons.

*Science as Manifested in Scientific Papers*
Morris and Yen (2004) conceptualize and interlink basic units of science using an entity-relationship model. Starting with 'collections of journal papers', they identify entity types such as papers, paper authors, references, and paper journals, see boxes in Figure 1. These entity types have also been called 'units of analysis' (Börner et al., 2003). The authors then add relations between different types of entities indicating direction and type of linkage. For example, each paper is associated with the author(s) who wrote it, the references it cites, the journal in which it was published, and the terms that appear in it.

Each box contains information on how the different entity types are used to derive insight. For example, the 'Papers' box is associated with information about:

- Co-authorship, i.e., author co-occurrence, groups oeuvres,
- Bibliographic coupling, i.e., reference co-occurrence, used to identify research frontiers,
- Paper co-occurrence in journals resulting in groupings of papers.
- Term co-occurrence, used to identify topically related papers.

Known laws can also be placed on the diagram. For example, Zipf's Law (1949) is placed between the 'Term' and the 'Papers' box as it describes the frequency of occurrences of terms in a text/paper as $rf=c$, where $r$ is the rank of a word, $f$ is its frequency, and $c$ is a constant. Bradford's Law (1934) can be seen between the 'Papers' and 'Journals' box as it characterizes the scattering of papers over different journals/fields. Lotka's inverse square law (1926) states that the number of authors producing $n$ papers is proportional to $1/n^2$. It is appropriately positioned between the 'Papers' and 'Authors' boxes.

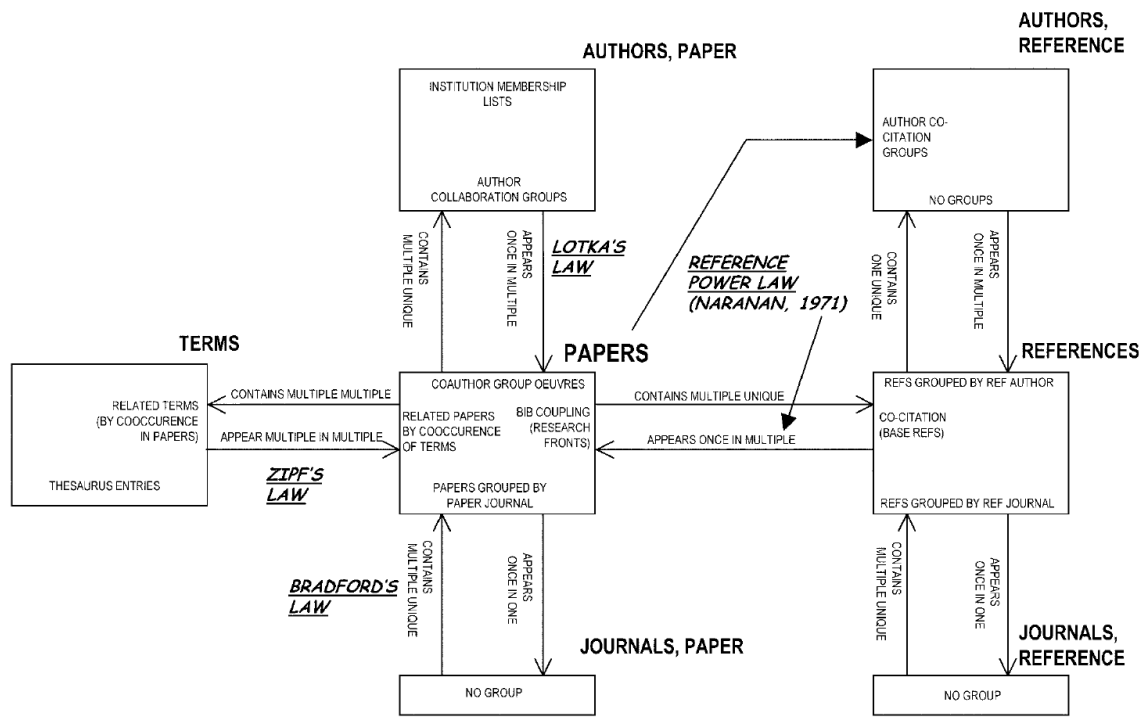

**Figure 1:** Diagram of the entities within a collection of journal papers, their directed links to each other, and co-occurrence groups, taken from (Morris & Yen, 2004).

In this conceptualization, papers and references are considered distinct entity types – not as roles a paper might have. The authors argue that this separation is necessary because papers and references represent different concepts and support different insights. For example, a paper is typically seen as a research report, whereas a reference represents a symbol of knowledge (Small, 1978). The same applies to paper journals, reference journals, paper authors, and reference authors as separate entity types. This conceptualization describes collections of papers exclusively. Funding, patents, books etc. are not captured.

*Science as Accumulation of Knowledge*

Recent work in mapping science (Börner et al., 2003; Shiffrin & Börner, 2004) has lead to locally and globally accurate maps of science (Klavans & Boyack, 2006a, 2006b) with support the assignment of 'science locations' to scholarly entities (e.g., authors, papers, patents, grants) based on journal names or keywords.

For example, the 'UCSD Basemap of Science' was created by Boyack and Klavans (Klavans & Boyack, 2007) based on 7.2 million papers and over 16,000 separate journals, proceedings, and series from Thomson Reuters' Web of Science and Elsevier's Scopus database over a five year period, 2001-2005. Bibliographic coupling using both highly cited references and keywords was applied to determine the similarity of journals. The final layout step was done using the 3D Fruchterman-Rheingold algorithm in Pajek; the results were so close to spherical (i.e., no nodes were in the middle) that all nodes were given a unit distance from the 'center of mass', resulting in a spherical layout. To ease navigation and exploration, a Mercator projection was applied to convert the spherical layout into the 2-dimensional map, see Fig. 2. Dots represent groups of topically similar journals. Links denote strong bibliographic coupling relations. Major areas of science are color coded and labeled.

As every node on the map represents a set of journals, scholarly entities can be overlaid based on matching of journal names. Fig. 2 shows the UCSD Basemap of Science with an overlay of papers relevant for the mapping of knowledge domains as identified in (Börner et al., 2003)[5]. Node size represents the number of papers per node. Each node also has an extensive list of key phrases as metadata that can be used to 'science-locate' non-journal entities (analogous to the use of longitude and latitude information to 'geo-locate' objects on a cartographic map). This map and variations on it are used to overlay other datasets, e.g., the core competencies of institutions or countries, comparisons of funding[6] vs. research results, or animations of topic growths over time.

---

[5] Note that six journals are not shown in Fig 2 as the base map provides no information on them. These are *Zentralblatt für Bibliothekswesen*, *Current Contents*, *Library and Information Science*, *Nachrichten für Dokumentation*, *International Forum on Information and Documentation* and *Proceedings of the American Society for Information Science*.
[6] Funding overlays for U.S. agencies on the UCSD map of science are shown at http://www.mapofscience.com.

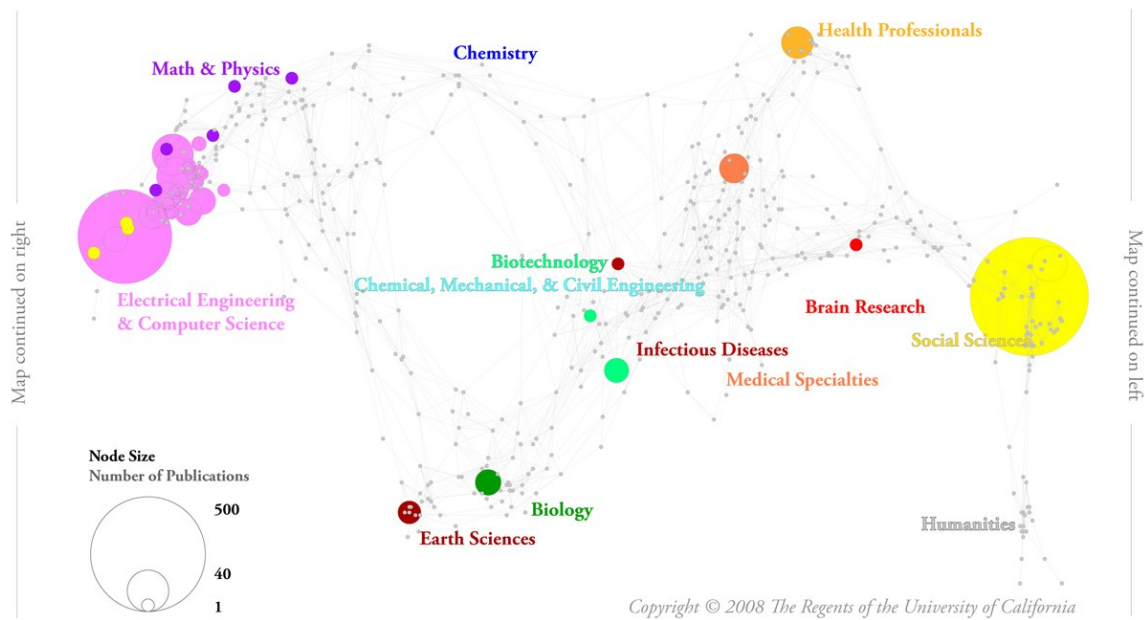

**Figure 2:** UCSD Base Map of Science by Richard Klavans and Kevin W. Boyack with overlay of publications relevant for the mapping of science.

The UCSD map shows the current structure of science. However, how did it evolve? How does it relate to major inputs of science such as people or funding? Daniel Zeller's "Hypothetical Model of the Evolution and Structure of Science" (see Figure 3) attempts to address these questions. The hand drawing conceptualizes science as interconnected layers of immortal knowledge generated by mortal authors. Starting with the very first scientific thought, science grows outwards in all directions. Each year, another layer is added to the 'global brain' shaped manifestation of knowledge (Börner, Dall'Asta, Ke, & Vespignani, 2005). New fields emerge (blue), established fields (brown) merge, split, or die. The cut-out reveals a layering of fat years with ample funding and industry buy-in that produce many new papers and slim years due to wars and economic collapses in which few papers are added. Each research field corresponds to a tube-shaped object. Some fields have very fast growth patterns due to electronic papers that interlink within days. Other fields communicate knowledge via books – years might pass before the first citation link is established. Blue tentacles symbolize the search for opportunities and resources and activity bursts due to hype and trends. The injection of money (yellow) has a major impact on how science grows. There are voids in our knowledge that might take the shape of monsters. Knowledge that is falsified or rendered obsolete will receive fewer linkages over time making it harder and harder to find it. Interconnections between existing and new knowledge islands strengthen and weaken over time. The trajectories of scientists that consume money, write papers, interlink papers via citation bridges, and fight battles on the front lines of research could be overlaid. Yet, scientists are mortal. All they leave behind are the knowledge structures that future generations can use and build future knowledge on.

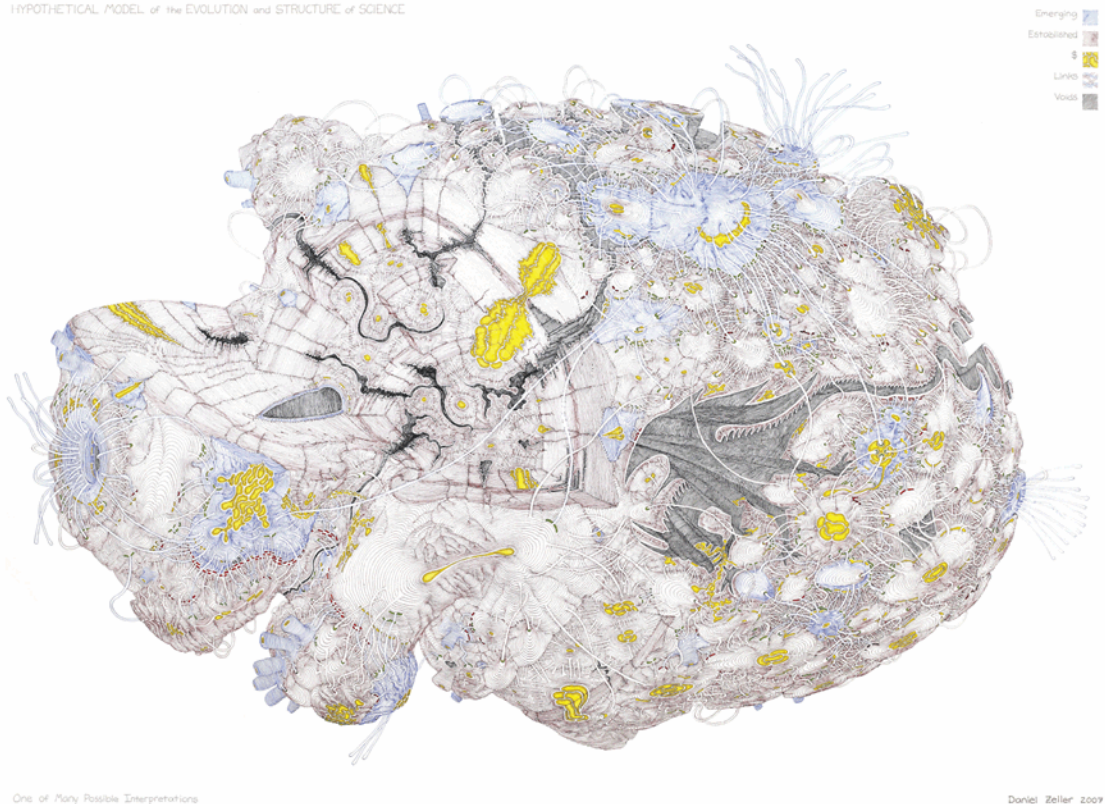

**Figure 3:** "Hypothetical Model of the Evolution and Structure of Science" by Daniel Zeller, 2007

The UCSD Base Map of Science shown in Fig. 2 shows the structure of the surface of the "Hypothetical Model". One day, digital datasets and compute power might exist to compute not only the surface but all interconnected layers of our scholarly knowledge in an automatic fashion.

Using this conceptualization, one can now science-locate the theories and results from different sciences.

On a *temporal scale*, historians of science are able to start at the very beginning of scientific thought. Visualizations of trees of scientific thoughts (Collins, 1998) or scholarly genealogies of scientific schools (e.g., the Stammbaum of Justus Liebig in von Dechend (1953), reproduced in (Bruckner & Scharnhorst, 1986)) can be shown as integral part of this science organism. Bibliometric/scientometric studies based on journal data could go as far back as 1665 when the first papers were published by the Royal Society.

*Topic wise*, research can be science located by topical coverage of a study ranging from small, often ethnographic, studies of a specific research community or body of literature, via medium size studies that might explore all of physics or more interdisciplinary areas such as nanoscience or neuroscience, to studies that aim to capture all of the sciences. The combination of a temporal dimension (from inside to outside) and topical dimensions (distance or similarity of disciplines) supports a visualization of the growth rates of fields. Historiographs, evolving co-authorship networks, invisible colleges, scientific schools, case studies, and individual trajectories of researchers can be located in this complex imagination.

Of course, most studies have a *geographical focus* as well, e.g., just US or only English material, and could be highlighted in the topical reference system or overlaid on a map of the world.

*Science as Search in an Abstract Problem Space*

The third conceptualization attempts to capture the growth, merging, and splitting processes in science. It conceptualizes science as search in a high-dimensional abstract landscape of problems (LP), see Fig. 4. In this LP, scientific fields represent known facts or actively researched areas while empty spaces represent unoccupied/unknown areas of science. Scientific fields are defined by scholars and their publications. In this "phenotypic space" of LP, scientific fields can grow, merge, and split. We can differentiate between incremental changes such as an extension of boundaries of existing fields or radical change of the LP due to 'scientific revolutions' (Kuhn, 1962). There is an explicit analogy to spatial models of evolutionary biology (Pigliucci & Kaplan, 2006) and many of the mathematical models from evolutionary biology and non-linear physics can be applied (Bruckner, Ebeling, & Scharnhorst, 1990; Ebeling & Scharnhorst, 2000; Scharnhorst, 2001). The LP model assumes the existence of a slowly changing fitness landscape (FL) and a much faster changing occupation landscape (OL). The FL -- also called evaluation landscape – is an expression of the actual attraction of scientific fields. The OL denotes the distribution of researchers over the different scientific fields. In this and in the previous conceptualization, the growth of a scientific field depends on those entering it as newcomers or experts. Newcomers, continuants, and terminants (Price & Gürsey, 1976) represent inflows, remaining population, and outflows. In a complex interplay between "random" search (mutations) and search based on comparison (selection and imitation) the OL dynamics unfolds as aggregation of many individuals. In the individual search process for a certain problem we find cognitive elements (i.e., which problem is interesting and can be solved?), social elements (collaboration), institutional boundaries (funding schemes, research agendas), as well as personality factors. It is obvious that the FL is shaped while being searched by activities in the OL.

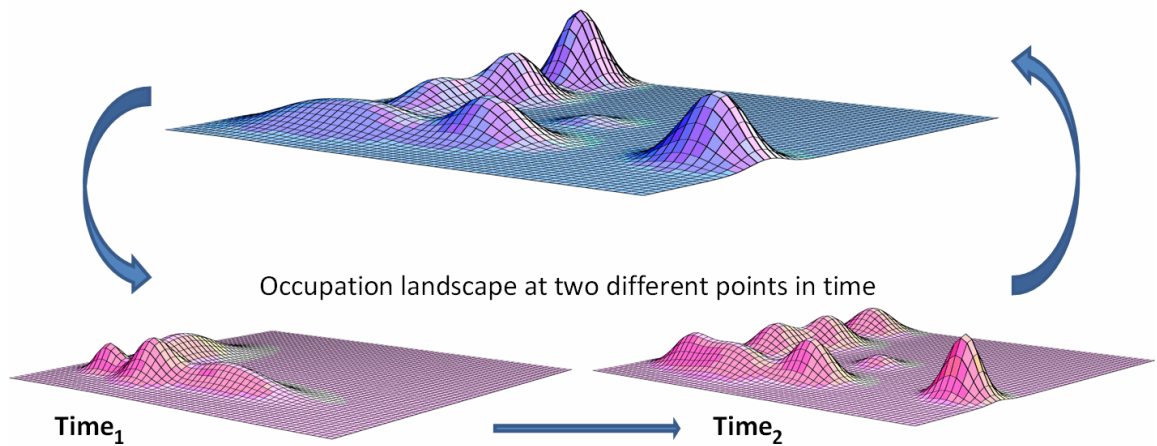

Evaluation or fitness landscape which can change over time with the occupation

Occupation landscape at two different points in time

Time$_1$ → Time$_2$

**Figure 4:** Abstract fitness landscape and possible occupation landscapes at two different time points

The images in Fig. 5 show the application of this conceptualization to the search of countries for high scientific profiles. Shown are the occupation spaces of 44 countries for two time periods. The two dimensional plots use percentages of a countries' total publications in the life sciences (L) on the x-axis and in physics (P) on the y-axis. Each country is represented by a dot and labeled by three letters, e.g., DNK is Denmark, PRC is People's Republic of China. From the first to the second time period, highly industrialized countries such as USA, Israel, and The Netherlands merge to a more coherent group of life sciences while previously isolated countries such as China, Russia, and most of the East-European countries become even more isolated in the second half of the 1990s than they had been before.

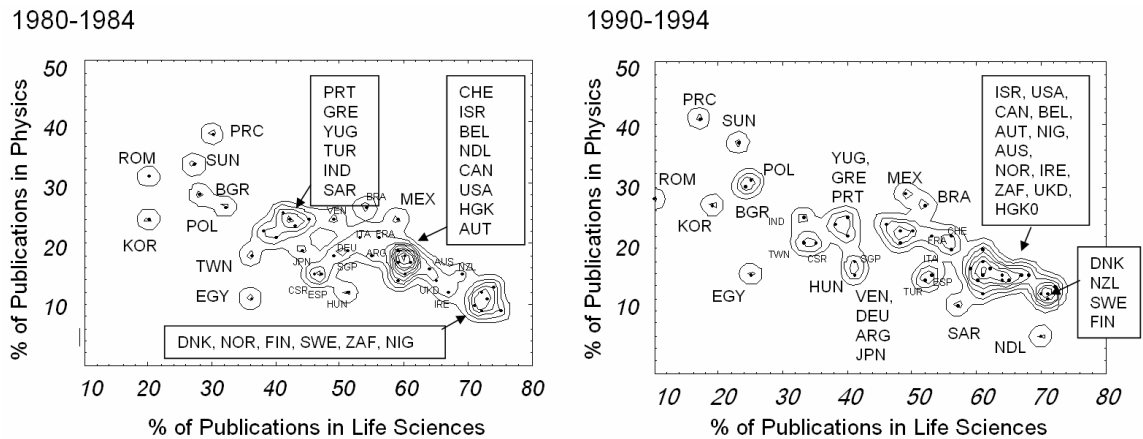

**Figure 5:** Data-based life-science vs. physics publications occupation landscapes for 44 countries

Comparing this conceptualization to the previous two, the OL can be represented either with AUTHORS (or agglomerations of authors) contributing to a specific problem/topic, or with PAPERS, or with TOPICS (see box in Fig.1). In each of these cases, the space in which elementary units are represented may also differ. It can be a space which coordinates are formed by all documents or by a classification as in the example above. The choice of unit and space will also influence how the result can be interpreted. But, in any case the FL resembles an evaluative dimension, probably best represented as function of REFERENCES. Authors interested to increase their reputation are expected to publish on 'hot topics', cite high quality works, and publish in high impact journals, etc.

Compared to 'Science as Accumulation of Knowledge', the high dimensional FL has resemblance with science maps such as the one shown exemplarily in Fig. 2 or the surface of the science object given in Fig. 3. Classification systems developed in library sciences, e.g., the visionary work of Paul Otlet (van den Heuvel & Rayward, July 11, 2005) or Salton's *vector-space model* (Salton, Wong, & Yang, 1975) are valid alternatives.

The selection of 'the best' conceptualization will depend on the scientific hypothesis one attempts to validate or the practical questions that require an answer. Ideally, visual conceptualizations of science help create a "trading zone" (Galison, 1997) in which definitions, approaches, and algorithms developed in different contexts and sciences can be compared and combined and differences can be "sorted out". The latter has been shown to be a powerful activity in knowledge production by (Bowker & Star, 1999). However, to sort out differences between conceptualizations, these must be made visible and explicit and this is what the authors in this issue aim at. Eventually, this process might lead to a typology of conceptualizations of science along dimensions such as scalability (from micro to macro phenomena), complexity (detailed complex models vs. generic models of basic complex mechanisms), applications (literature based discovery to science policy), degree of operationalisation, etc.

**Contributions of Papers**

This special issue invited contributions on topics such as reviews of existing conceptualizations of the structure and evolution of science. Philosophers, sociologists and historians of science, scientometricians, and other authors were encouraged to compare and contrast multiple works; historiographic and ethnographic work on how people understand and communicate the structure and dynamics of science via imagery and textual descriptions. Papers in this category were supposed to analyze a variety of approaches, including critiques on science conceptualizations; and novel conceptualizations and empirically validated models of science and scientific communication in which authors discuss epistemic assumptions and disciplinary roots, possible application domains, covered and omitted features of scientific evolution, and model interpretation.

In the first round of abstract submissions we received 37 abstracts from 16 countries and 16 areas of science comprising Bibliometrics, Biology, Biomedical Informatics, Ecology, Education, Geography, History of Science, Informetrics, Molecular Biology, Philosophy of Science, Science Semantics, Scientometrics, Semantic Analysis, Socio-Futurism, Sociology, and Sociophysics. The number and diversity of submissions might serve as another indicator for need and timeliness of the topics addressed by this issue. Each abstract was reviewed regarding its match with the above topics. Highest scoring author teams were invited to submit full papers. Submitted full-papers were reviewed and the best papers included in this issue. Subsequently, we group the eight papers, highlight their main contributions and discuss their interrelations.

*Historical Context*

As discussed in section 'Science of Science History', a science of science conceptualization has to synergistically draw on existing works from many different fields of science.

> **Eugene Garfield** in "*From The Science of Science to Scientometrics Visualizing the History of Science with HistCite Software*" traces the history of quantitative studies of science back to the early 20[th], discusses pioneers, major publications, and their interconnections in the evolving field.

He uses his HistCite software to visualize the impact of Price's works on the growth of the field based on a ranked citation index of the 100,000 references cited in the 3,000 papers citing Price.

*Collaboration and Communication Networks*

Science can be seen as an evolving system of different basic units of science that are tightly linked and dynamically coupled. Networks are just one 'product', one 'trace', or one 'representation'[7] of the collective, self-organized emerging structures in science. But they allow the linking of structural properties to dynamic processes. 'Structural holes' seem to be relevant both in collaboration and in citation networks. Chen et al. present an explanatory theory of scientific discovery based on an extended theory of structural holes and conclude that the nature of transformative discoveries can be characterized by structural and temporal properties of forging a path spanning over a structural hole in a conceptual world. Similarly, Lambiotte and Panzaraza argue that scientists at the boundaries of established, well-connected communities can be crucial for the spreading of new ideas. Obviously, a balance exists between structure formation (visible in modularity and community formation) and structures destruction (e.g., caused by scientific revolutions). Bettencourt et al. show how the topology of collaboration networks changes over time. Topological transitions can be related to phases of stabilization and settlement as well as to phases of new discoveries. Also, Chen et al discuss that scientific discoveries, the emergence of new journals, and new research areas cause systematic transitions in the network structures. Network analysis and modeling provides an alternative framework to order and systematize already known insights in the psychology, sociology and history of scientific change and discoveries. The embedding of network approaches in social science theories poses an additional challenge.

> **Renaud Lambiotte and Petro Panzarasa** contributed *"Communities, Knowledge Creation and Information Diffusion"*. The authors reflect upon optimal positions of the scientist in collaborative networks concerning information reception and use concepts of social network analysis as social capital or weak ties. Extending beyond the individual perspective the authors discuss advantages and disadvantages of close scientific communities and sparsely connected ones concerning information diffusion. They also point to differences between information diffusion and other diffusion processes and the need to network models for multimodal and modular networks.

> **Chaomei Chen, Yue Chen, Mark Horowitz, Haiyan Hou, Zeyuan Liu and Donald Pellegrino** contributed *"Towards an Explanatory and Computational Theory of Scientific Discovery"*. The authors present an explanatory theory that conceptualizes scientific discoveries as a brokerage process and also unifies knowledge diffusion as an integral part of a collective information foraging process. As suggested in the call for papers, their model interlinks different works on scientific growth and knowledge diffusion.

---

[7] Personal communication Iina Hellsten.

**Luis Bettencourt, David Kaiser and Jasleen Kaur's** paper is entitled *"Scientific Discovery and Topological Transitions in Collaboration Networks"*. Bettencourt and co-authors look at topological changes in the life cycle of collaboration networks. They find that following a period of scientific discovery, collaboration networks become more densely interconnected, as shared concepts and tools knit practitioners together. The dynamics of these topological transitions show remarkable similarity across a broad range of scientific fields.

*Spatial Concepts*

Cartographic maps of our Earth are the outcome of a long struggle for the right unit of analysis, measurement techniques, visual representations, and interpretations (Wilford, 1981). Maps of science will have to undergo a similar normalization and standardization process yet should draw on and learn from the rich theory and practice in cartography. The contributions of Skupin and Frenken et al. point in promising directions but also warn for potential pitfalls. As discussed by Frenken et al., geography is inherent to scientometrics studies particularly as they concern national scientific gains—such comparisons are ever more important in a globalised world. Suddenly the geographic differences turn from an object of gross performance into a mechanism of understanding knowledge production as social, cognitive, and cultural interaction. Logically, Frenken et al embed their theory of geographic proximity into a wider theory of the proximity of knowledge processes in very different dimensions—social, organizational but also cognitive. These are the same dimensions addressed in the sociology of knowledge of Schroeder & Meyer. Skupin posits not only how to measure these 'spatialities' but how to map them as well. He shows that the choice between discrete or continuous representations of units of analysis has deeper implications than mere convenience in terms of computational algorithms.

**Koen Frenken, Sjoerd Hardeman and Jarno Hoekman** in *"Spatial Scientometrics: Towards a Cumulative Research Program"* discusses a "geographical turn" in scientometrics pointing to the relevance of the local in a globalised world where distance has supposedly disappeared.

**André Skupin** in *"Conceptualizing Science: Implications for Knowledge Domain Visualizations"* shows how different models of spatiality influence the visibility of scientific change on the level of the individual scientist as well as on the level of fields. Skupin argues that a study of science should not be driven by available data (formats) and algorithms but by a theoretical conceptualization of science that addresses/matches the insight needs of different stakeholders, e.g., historians of science, science policy makers, or children trying to make sense of science. The author poses the question if bibliometric entities are merely situated in the space of science or if they make up the space of science arguing that growth *within* a space is very different from growth *of* that space.

*Re-Conceptualizations*

Critical reflection of current theories in the social studies of sciences are based on sociological theories of scientific change such as Kuhn's theory of scientific revolution (Kuhn, 1962), Luhmann's sociology of communication (Luhmann, 1990) or Witley's link between the cognitive and social organization of science (Whitley, 1984). The papers of Lucio-Arias & Leydesdorff and Schroeder & Meyer depart from a critical reflection of current theories in the social studies of sciences to head in different directions. While Lucio & Leydesdorff elaborate on a close, coherent theory of scientific communication, the contribution of Schroeder & Meyer advocate for a sociological approach which broadens the sociology of science towards a sociology of knowledge (Schroeder, 2007) For Schroeder & Meyer the prematurity and the diffuse character of a new research area allows an opportunity to look in parallel at cognitive, communicative, institutional and organization structures to identify "critical events". Lucio & Leydesdorff present a stringent explanation of scientific change as visible in the system of scientific journals. The seeming reductionism to one "language" is treated against the possibility to formulate and to test a hypothesis.

**Ralph Schroeder and Eric T. Meyer** in *"Untangling the Web of e-Research: Towards a Sociology of Online Knowledge"* provides arguments for why a reconceptualization of science is needed in the light of increasing team efforts, interdisciplinarity, technology, and online knowledge creation. Taking e-research as a provocative case, the authors discuss the need of a sociological approach to knowledge -- including science. Combining quantitative and qualitative methods they seek for operationalizations of a sociological approach which includes the actors, the instructions, their location and the technological research environment they are embedded in.

**Diana Lucio-Arias and Loet Leydesdorff** in *"The Intellectual Self-organization of Scientific Knowledge and the Literary Model of Scientific Communication"* develop a communication theory of autonomous and self-organized scientific change. Starting with scientific publication as a basic unit of analysis and referencing as an elementary process they argue in favor of a stage-like process of increasing and decreasing uncertainty. Uncertainty can be measured in communications by looking at word use or changing composition of journal citation networks. With their approach they shape empirical evidence for the interplay between codification and structuring (paradigm setting) and the breaking up from these structures towards new interdisciplinary adventures -- an interplay of constructive and destructive factors shaping the trajectory of scientific knowledge.

**Concluding Remarks**

The conceptualizations and models of science presented in this issue differ considerably in the
- Scope of the modeled system or object (from very specific fields of science to all of science but also from very specific static or dynamic effects to major interlinked effects),
- Basic building blocks of science (from papers, authors to scientific fields, institutions, regions),
- Interactions of building blocks (via papers references and co-author linkages to spatial and topical distributions),
- Basic mechanisms of growth and change captured (indicators, network evolution, trajectories in geographic and topic spaces),

- Existing laws encapsulated (static and/or dynamic).

Eight papers and an editorial cannot possibly capture the breath of relevant work and the number of puzzle pieces needed for a theoretically grounded and practically useful science of science. Instead, they should be seen as 'stepping stones' towards the envisioned shared conceptualization of science. In this process visual conceptualizations of the structure and dynamics of science have an important role to play as they help diffuse and combine existing and new knowledge and expertise across disciplinary, cultural and geospatial boundaries.

**Acknowledgements**

This editorial did benefit from expert input provided by Blaise Cronin, Kevin W. Boyack, Margaret Clements, Manfred Bonitz, Paul Wouters, Anne Beaulieu, and authors of papers in this special issue. We would like to thank Leo Egghe (Editor of the Journal of Informetrics) for his support of this special issue. Mark A. Price, Indiana University served as an editorial assistant and his much appreciated patience and diligence was instrumental in the organization of the review process.

This work was partially funded by the National Science Foundation under Grant No. SBE-0738111 and a James S. McDonnell Foundation grant in area Studying Complex Systems. Any opinions, findings, and conclusions or recommendations expressed in this material are those of the author(s) and do not necessarily reflect the views of the National Science Foundation.